\documentclass[aps, pop, amsmath,amssymb, reprint,onecolumn,superscriptaddress,showkeys,showpacs]{revtex4-1}

\include{bookdefs}

\usepackage{color}
\usepackage{rotating}
\usepackage{amsmath,amsthm,amsfonts,amscd}
\usepackage{graphicx}
\usepackage{xspace}
\usepackage{amsmath}
\usepackage{placeins}

\renewcommand{\vec}[1]{\mbox{\boldmath$#1$}}

\definecolor{orange}{rgb}{.9,.3,0}

\newcommand{\jorek}{\texttt{JOREK}\xspace}

\usepackage{float}
\usepackage{calc}

\begin{document}

\title{Early evolution of electron cyclotron driven current during suppression of tearing modes in a circular tokamak}

\author{J. Pratt}
\email[]{j.l.pratt@exeter.ac.uk}
\affiliation{Astrophysics Group, University of Exeter, Exeter, EX4 4QL, United Kingdom}
\affiliation{FOM Institute DIFFER, Dutch Institute for fundamental Energy Research,  \\
 Eindhoven, the Netherlands}                                 
\author{G.T.A. Huijsmans}
\affiliation{CEA, IRFM, 13108 Saint-Paul-Lez-Durance, France}                                 
\author{E. Westerhof}
\affiliation{FOM Institute DIFFER, Dutch Institute for fundamental Energy Research,  \\
 Eindhoven, the Netherlands}                                 

\date{\today}

\begin{abstract}
When electron cyclotron (EC) driven current is first applied to the inside of a magnetic island, the current spreads throughout the island and after a short period achieves a steady level.
Using a two equation fluid model for the EC current that allows us to examine this early evolution in detail, we analyze high-resolution simulations of a 2/1 classical tearing mode in a low-beta large aspect-ratio circular tokamak.  These simulations use a nonlinear 3D reduced-MHD fluid model and the \jorek code.  
During the initial period where the EC driven current grows and spreads throughout the magnetic island, it is not a function of the magnetic flux.  However, once it has reached
a steady-state, it should be a flux function.  We demonstrate numerically that if sufficiently resolved toroidally, the steady-state EC driven current becomes approximately a flux function.  We discuss the physics of this early period of EC evolution and its impact on the size of the magnetic island.
\end{abstract}


\maketitle

\vspace{5mm}

\section{Introduction}

Above a critical value of the plasma beta, neoclassical tearing modes (NTMs) are destabilized forming magnetic islands.  As the size of magnetic islands grow, hot particles are more easily lost from the machine and disruptions can be triggered.  Thus the formation of magnetic islands establishes an effective limit on the plasma beta that can be achieved in a fusion device.  The presence of even small magnetic islands has a nonlocal effect on a fusion plasma.  Magnetic islands change a plasma's temperature profile, the current profile, and the rotation velocity \citep{waelbroeck2009theory}.   The primary tactic for controlling and suppressing NTMs is to apply electron cyclotron current drive (ECCD) in a localized area inside the magnetic islands \citep{isayama2005steady, petty2004complete, gantenbein2000complete,la2006neoclassical,isayama2000complete,maraschek2012control}.

In the limit of low plasma beta, the neoclassical bootstrap current, which distinguishes a neoclassical tearing mode from a classical tearing mode, is negligibly small.  The low beta limit represents a simplification of the full physics of fusion plasmas, which is still of interest \citep{hazeltine1975kinetic,chu2002study,coelho2007effect}.  In this work we study the dynamics of the electron cyclotron driven current within magnetic islands produced by a classical tearing mode in the low beta regime. 

Experimentalists can reconstruct the width and position of a magnetic island in a fusion plasma from 
measurements of the electron temperature obtained from electron cyclotron emission; this profile flattens inside a magnetic island. Changes in the island width can also be calculated from the magnetic field perturbation, which is constructed from an integrated and weighted sum of Mirnov coil measurements \citep{maraschek2012control,urso2010asdex}.  However experiments have not been able to produce a complete picture of the dynamics internal to the island during stabilization.  A deeper understanding of this physical process is necessary in order ultimately to build a more detailed theory of stabilization than the generalized Rutherford equation provides.

Approaching the growth and suppression of tearing modes with simulation allows us to observe how physical processes inside a magnetic island develop on time and space scales finer than those described by the generalized Rutherford equation, or measurable in experiment.  Over the last few decades computing power has increased to a level that, with a magnetohydrodynamic (MHD) fluid model, the nonlinear
development of instabilities are routinely simulated with sufficient resolution in both space and time to produce results with excellent accuracy.
Several large codes have been developed to solve different variations of the MHD fluid equations in a tokamak.  These codes include
 NIMROD \citep{sovinec2003nimrod,sovinec2005nonlinear}, BOUT++ \citep{dudson2009bout++,xia2013six}, M3D(-C1) \citep{sugiyama2001studies,breslau2011onset}, XTOR(-2F) \citep{lutjens2010xtor,halpern2011diamagnetic}, and \jorek \citep{huysmans2007mhd,czarny2008bezier}.
The \jorek code has been used to simulate MHD instabilities in tokamaks successfully
 using several nonlinear reduced MHD models \citep{huijsmans2013non,starwall} as well as full MHD models \citep{haverkort2013magnetohydrodynamic,Haverkort2016}.  In this work we use \jorek and a reduced MHD fluid model to simulate the evolution of classical 2/1 tearing modes from their first development, through their saturation and eventual partial suppression using electron cyclotron current drive (ECCD).  

We have recently revisited the closure of single fluid MHD in the presence of ECCD \citep{westerhof2014closure}.  A typical feature of ECCD is the two step process characterizing the current drive described by Fisch and Boozer \citep{fisch1980creating,fisch1987theory}: the electron cyclotron waves create an asymmetry in the collisionality of the electron distribution which then results in the creation of a net current with negligible momentum transfer between electrons and ions. The effect of this current is expressed as a commonly-used modification to Ohm's law, which is closed by an equation for the evolution of the EC driven current, reflecting the delayed nature of its source and its convection with the parallel velocity of resonant electrons.
Previous modeling efforts have used a single equation for the EC driven current \citep{giruzzi,hegna2009}.  Our two-equation fluid closure for the EC driven current therefore results in a model that is more accurate, particularly for the early evolution of the current.

We investigate the physical consequences of this two-equation fluid closure for the EC current presented in \citet{westerhof2014closure}.  This work is organized as follows.  Section \ref{secmodels} summarizes the physical model implemented for EC current and the magnetohydrodynamic fluid model and numerical models used in the \jorek code.  Section \ref{secsims} describes the simulations of 2/1 tearing modes performed with \jorek.  Section \ref{secresults} presents our numerical results for the early dynamics of the EC driven current.  In Section \ref{secconcl} we discuss implications of these results.

\section{Physical and Numerical Models \label{secmodels}}

\subsection{Physical model for electron cyclotron current}

We employ the fluid closure described in \citet{westerhof2014closure}, which has been designed to model the Fisch-Boozer current-generation mechanism \citep{fisch1980creating,fisch1987theory}.  Using this current-generation mechanism, the electron cyclotron waves drive an excess of electrons at the resonant parallel velocity and at high perpendicular velocity.  There is a corresponding shortfall of electrons at the resonant parallel velocity and at low perpendicular velocities.
These two populations of electrons exhibit different collisionality, allowing a steady-state current to emerge.
The source region of this current cannot be equated with the region where EC power is deposited, but is extended along the magnetic field lines crossing through the EC power deposition region.  The current is then convected out of the region where it is generated.  Thus the essence of our closure model are \emph{two} current equations that describe the convection of the applied EC current along the magnetic field lines:
\begin{eqnarray}\label{eqfluidclosure1}
    {\partial j_1 \over \partial t} = - S_{\mathrm{ec}} - \nu_1 j_1 + v_{\parallel, \rm res} \nabla_\parallel j_1 ~,
\\ \label{eqfluidclosure2}
    {\partial j_2 \over \partial t} = + S_{\mathrm{ec}} - \nu_2 j_2 + v_{\parallel, \rm res} \nabla_\parallel j_2 ~.
\end{eqnarray}
In eqs.~\eqref{eqfluidclosure1} and \eqref{eqfluidclosure2} the parallel velocity of the electrons resonant with the electron cyclotron waves is given by $v_{\parallel, \rm res}$, and $\nabla_\parallel$ is the gradient parallel to the magnetic field.   The current perturbations $j_1$ and $j_2$  each have a source $S_{\mathrm{ec}}$, which is of equal strength but different sign for each current perturbation, corresponding to the amplitude of the current perturbation of the population of electrons that resonates with the EC wave. 
The source region of the EC driven current is extended over a toroidal length equal to $v_{\parallel, \rm res}/\nu_1$. For typical tokamak parameters this corresponds to $ \mathcal{O} (10^2)$ toroidal revolutions.  The regime of fast island rotation, where $\nu_{\mathsf{rotation}}/\nu_1   \gg 1$, is appropriate for the non-rotating tearing mode that we study in this work.  We thus adopt a form for the EC source $S_{\mathrm{ec}}$ that is only radially localized.  We model $S_{\mathrm{ec}}$ as a Gaussian distribution, centered at the resonant surface in the poloidal plane where the magnetic islands begin to grow.  The EC source has a constant amplitude and narrow standard deviation.

The different collision frequencies of the two populations of electrons that produce the current perturbations in eqs.~\eqref{eqfluidclosure1} and \eqref{eqfluidclosure2} are $\nu_1$ and $\nu_2$, respectively and $\nu_1>\nu_2$.
 The difference in $\nu_1$ and $\nu_2$ produces a difference in $j_1$ and $j_2$ as these current perturbations evolve.  The current perturbation $j_1$ decays quickly over the distance $v_{\parallel, \rm res}/\nu_1$, while $j_2$ decays more slowly.  This creates a net EC driven current  
\begin{eqnarray}\label{eqtotalcurrent}
    j_{\mathrm{ec}} = j_1+ j_2~.
\end{eqnarray}
which decays at the slower collision rate $\nu_2$.  After the EC source is applied, a short period of time is required for $j_{\mathrm{ec}}$ to reach a steady-state value.  It is this early period of ECCD application that we examine in this work.

 \subsection{A reduced MHD fluid model in the \jorek code}

We use the physical model for the EC current of eqs.~\eqref{eqfluidclosure1}-\eqref{eqtotalcurrent} in conjunction with a reduced MHD model, which is one of several MHD models implemented in the \jorek code.  \jorek is a toroidal code that has the capacity to accurately model the geometry and divertor, consistent with different tokamak designs.
In the reduced MHD model we employ, vector fields are represented in terms of the poloidal magnetic flux $\psi$, the velocity stream function $u$, and the parallel velocity $v_{\parallel}$.  The full magnetic and velocity fields can be reconstructed
from these functions using the definitions:
\begin{eqnarray} \label{bfielddef}
\vec{B} &=& - \hat{e}_{\phi} \times \frac{1}{R} \nabla \psi    + \hat{e}_{\phi} \vec{B}_{\phi} ~,
\\ 
\vec{v} &=&  \hat{e}_{\phi} \times R \nabla u  + v_{\parallel} \hat{b} ~.
\end{eqnarray}
Here $\hat{e}_{\phi}$ is a unit vector in the toroidal direction, and $\hat{b}$ is a unit vector in the direction of the magnetic field $\vec{B}$.  $R$ is the major radius of the tokamak.  The toroidal magnetic field $\vec{B}_{\phi} = F_0/R$ is held constant; when the toroidal magnetic field is large, the reduced MHD formalism is accurate \citep{zank1992equations,franck2014energy}.  The evolution equations \citep{hoelzl2012reduced} for the poloidal magnetic flux $\psi$, and the velocity $v$ are: 
\begin{eqnarray} \label{psieq}
\frac{ \partial \psi }{\partial t} + R \left[ u, \psi \right]
&=& -F_0 \frac{\partial u }{\partial \phi} + 
\eta (\vec{j} -\vec{j}_{\mathrm{BS},0}-\vec{E_0}/\eta  - \vec{j}_{\mathrm{ec}} ) ~,
\\
 \label{poloidal_momentum_equation}
\rho \frac{\partial \vec{v}}{\partial t} &=&
- \rho (\vec{v} \cdot \nabla)\vec{v}- \nabla \vec{p} + \vec{j} \times \vec{B} + \nu \nabla^2 \vec{v} ~,
\end{eqnarray}
where the Poisson bracket has been defined in the standard way as $\left[ u, \psi \right]=\partial_R u \partial_Z \psi - \partial_Z u \partial_R \psi$.  A resistivity and viscosity, given by $\eta$ and $\nu$, are constant input parameters in these equations; their values in our simulations will be discussed in Section~\ref{secresults}.  The equilibrium electric field is given by $\vec{E_0}$.  The equilibrium bootstrap current
$\vec{j}_{\mathrm{BS},0}$ is not evolved in this work, allowing classical tearing modes to be produced. In eq.~\eqref{psieq} the EC driven current $\vec{j}_{\mathrm{ec}}$ is defined by our fluid model in eq.~\eqref{eqtotalcurrent}.  Additional terms for drifts and two-fluid effects are available in the \jorek code, but are not used in this work.
\jorek solves the momentum equation~\eqref{poloidal_momentum_equation} in the form of two separate equations for the
parallel velocity and the toroidal vorticity $\omega = \nabla^2_{\mathrm{pol}} u$.

\jorek is a compressible MHD code and solves coupled evolution equations for the density $\rho$ and temperature $T$:
\begin{eqnarray} 
\frac{\partial \rho}{\partial t} &=&  - \nabla \cdot (\rho \vec{v})+ \nabla \cdot (D_{\perp} \nabla_{\perp} \rho) ~, \label{densityeq}
\\ \nonumber
\rho \frac{\partial T}{\partial t} &=&  - \rho \vec{v} \cdot \nabla T - (\kappa -1) p \nabla \cdot \vec{v}
+ \nabla \cdot (K_{\perp} \nabla_{\perp} T + K_{\parallel} \nabla_{\parallel} T ) ~. \label{temperature}
\end{eqnarray}
Here $K_{\parallel,\perp}$ are the parallel and perpendicular heat diffusivities, and $\kappa=5/3$ is the ratio of specific heats.   The perpendicular heat diffusivity is assumed to be small in this work.  The parallel heat diffusivity is temperature dependent.

The toroidal current density $j_{\phi}$, is also calculated at each time step from:
\begin{eqnarray} \label{toroidal current}
j_{\phi} &=& R^2 \nabla \cdot (R^{-2} \nabla \psi )  ~.
\label{vorticity}
\end{eqnarray}  
In our simulations using \jorek, the time integration is carried out using a fully implicit BDF1 (backward differentiation formula) scheme due to Gear \citep{gear1971numerical}.    The preconditioning we use is a variant of block-Jacobi preconditioning, performed on a reordered matrix \citep{huysmansimplementation,latu2012non,franck2014preconditioning}.  The spatial scheme is a finite
element method in the poloidal plane of the tokamak.  The finite elements are based on bicubic B\'ezier surfaces, a generalization of
cubic Hermite elements.  The advantage of this is that the finite elements are aligned with
the equilibrium magnetic flux surfaces \citep{czarny2008bezier}.  In the toroidal direction, a spectral method is used.
For further details of the numerical models in the current \jorek code, we refer to \citet{franck2014energy,holzl2012reduced}.

\section{Classical tearing mode simulations with JOREK \label{secsims}}

\begin{figure}
\resizebox{3.5in}{!}{\includegraphics{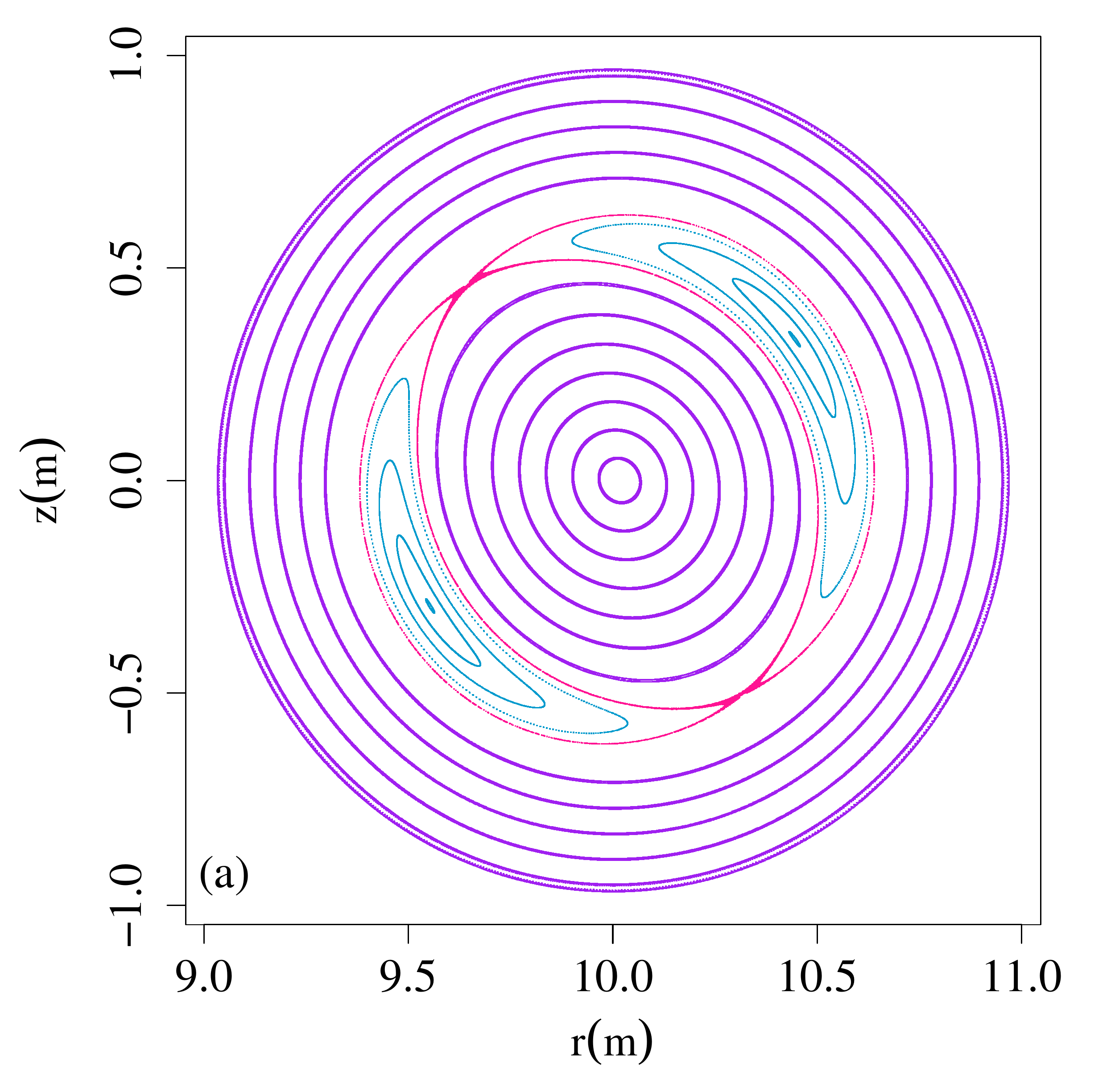}}\resizebox{3.5in}{!}{\includegraphics{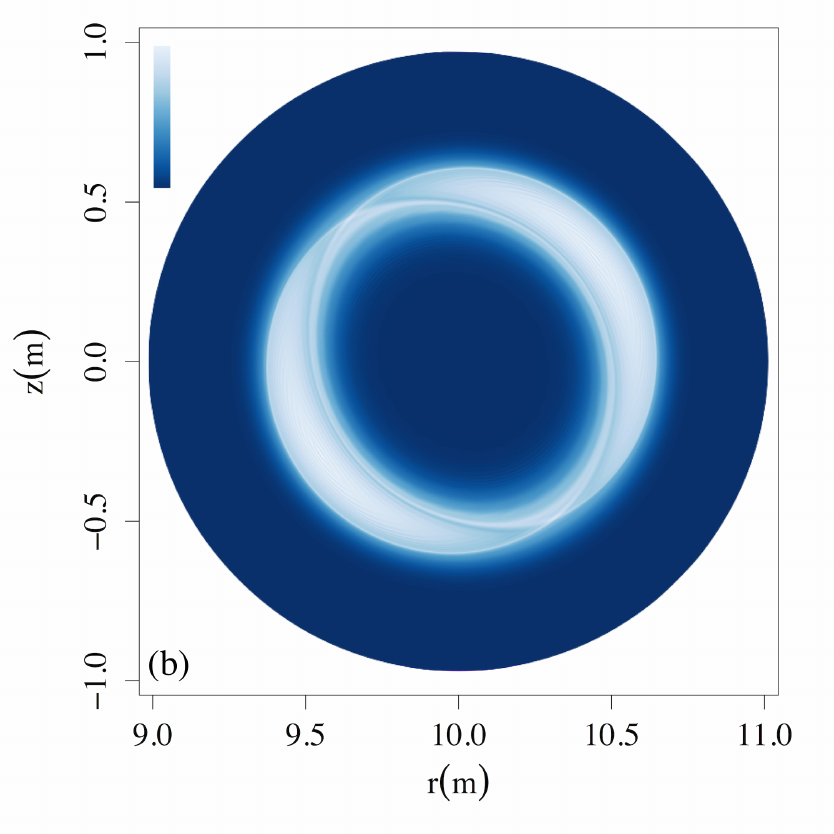}}
\caption{\label{figpoincare} A Poincar\'e map (a) of the poloidal plane of the circular tokamak in simulation T1 .  Shown in blue is the magnetic island formed by the 2/1 tearing mode at the point where they have come to saturation, and are at their largest width.  The separatrix is plotted in pink.  The nearly-concentric surfaces of constant magnetic flux outside the magnetic island are plotted in purple.   A visualization (b) of the EC driven current in the poloidal plane of the circular tokamak.  The EC driven current spreads along flux surfaces, filling the magnetic island in simulation F3.}
\end{figure}

We produce simulations of a classical $m/n=2/1$ tearing mode where $m$ and $n$ are the poloidal and toroidal mode numbers respectively.  This 2/1 classical tearing mode is produced in a circular tokamak with minor radius $a=1$m and major radius $R=10$m.  The maximum beta of the plasma is $\beta \sim  \mathcal{O} (10^{-4})$. 
The resonant surface is located at the point where the safety factor $q=2$; no other surfaces of rational $q$ are located inside the plasma, allowing us to study a pure 2/1 tearing mode that does not interact with other tearing instabilities.   A magnetic island is allowed to form from small fluctuations at the resonant surface at $q=2$.  We follow the growth of the magnetic island through the exponential growth phase until the size of the magnetic island reaches a steady large width, commonly referred to as saturation of the tearing instability.    Panel (a) of FIG.~\ref{figpoincare} shows a Poincar\'e map of the poloidal plane in our simulations.  In this figure, the magnetic island is shown at the point of saturation, when it has reached its largest size.   Panel (b) of FIG.~\ref{figpoincare} shows the outlines of the magnetic island in a visualization of the steady-state EC driven current.

\begin{figure}
\resizebox{3.5in}{!}{\includegraphics{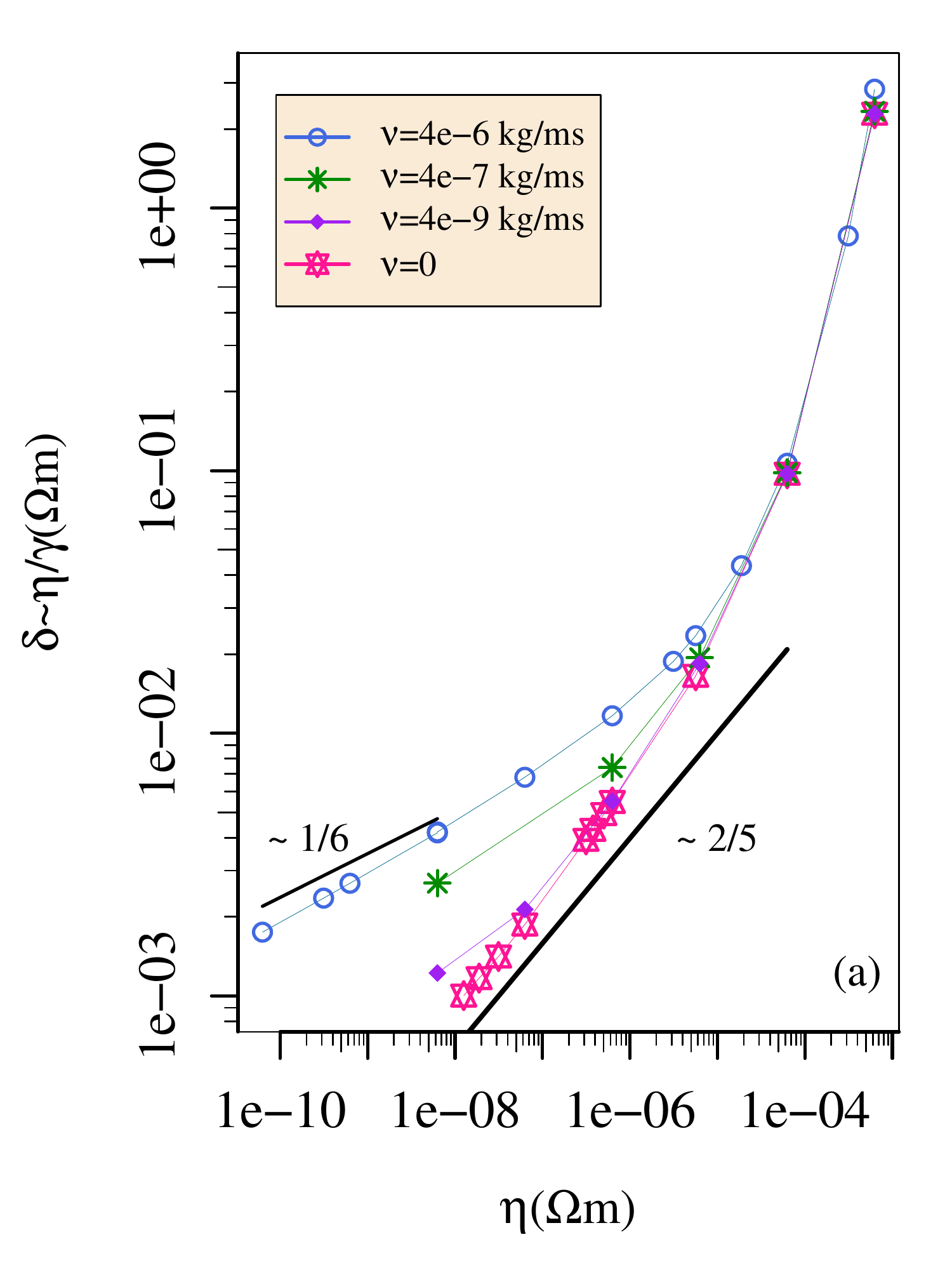}}
\resizebox{3.5in}{!}{\includegraphics{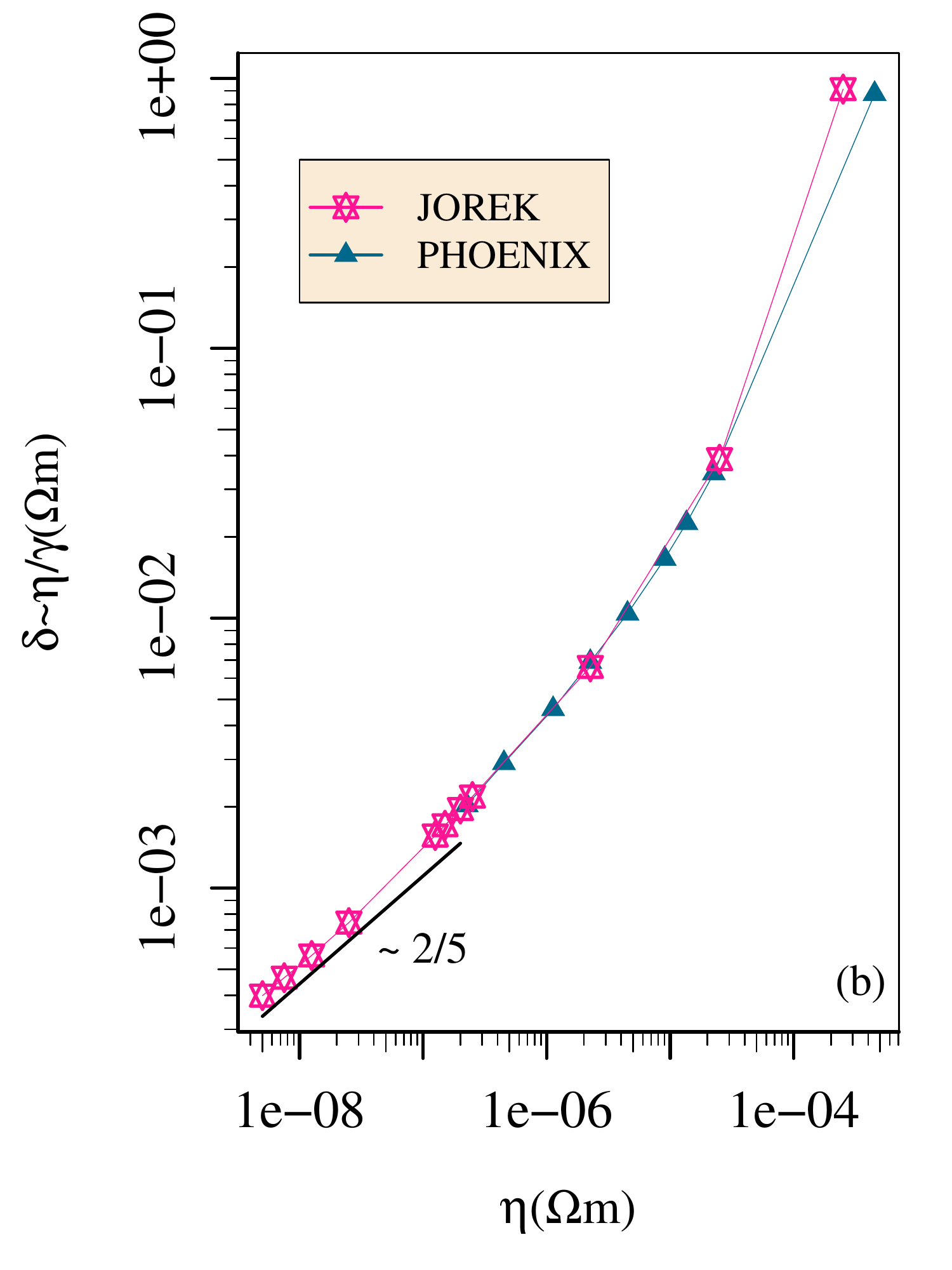}}
\caption{The width of the resistive layer $\delta$ vs the resistivity $\eta$ for (a) reduced-MHD \jorek simulations with different values of the viscosity $\nu$ and (b) identical simulations at viscosity $\nu=0$ performed with a reduced-MHD model in \jorek and with the linear stability code PHOENIX \citep{blokland2007phoenix} in the zero beta limit.  Heavy black lines show the theoretically predicted resistive scaling of $\eta^{2/5}$ and visco-resistive scaling of $\eta^{1/6}$.   \label{figbench} }
\end{figure}

A magnetic island forms in a narrow layer around the resonant surface where resistivity is important. This layer is commonly referred to as the resistive layer, or alternately has been referred to as the tearing layer, the singular layer, or the reconnection layer \citep[e.g.][]{wesson2011tokamaks,connor1988tearing,porcelli1987viscous}.  Following \citet[][]{porcelli1987viscous} eq.~(49), we note that the width of the resistive layer, $\delta$, scales as the ratio of the resistivity $\eta$ to the magnetic island growth rate $\gamma$ in the exponential growth phase. To confirm that our classical tearing mode simulations produce the expected results, we compare the width of the resistive layer with well-known scaling laws based on the resistivity and the viscosity.
Panel (a) of FIG. \ref{figbench} demonstrates that in \jorek simulations we recover the well-known $2/5$ scaling of the resistive layer width with resistivity \citep{biskamp:book} when viscous effects are negligible.  When resistivity is low and viscosity $\nu$ is high, the resistive layer width is expected to follow a theoretically determined viscoresistive scaling \citep{porcelli1987viscous} with $\eta$. \jorek simulations also produce this $1/6$ scaling with $\eta$ in the correct regime.    We observe that when resistivity is large, the resistive layer is also large, and this affects the formation of magnetic islands.  This increase of the resistive layer width is clearly shown in FIG. \ref{figbench}(a).   In this high resistivity regime, the theoretical scaling laws no longer apply.   

Our examination of the resistive layer width sheds light on the high resistivity regime.   At very high resistivity, the magnetic island growth rate no longer scales as $\gamma \sim \eta^{3/5}$; it displays a ``hook'' as resistivity is increased above $\eta \sim 10^{-5} \Omega $m, as shown in \citet{Haverkort2016}.  The resistive layer width, which rapidly increases above $\eta \sim 10^{-5} \Omega $m, provides a physical explanation for this reduced growth rate.  A large and diffuse resistive layer impacts magnetic reconnection and the early growth of the tearing instability.  The tearing stability parameter $\Delta'$ is calculated by matching of the ideal MHD linear solution outside of the resistive layer to the resistive MHD linear solution inside the resistive layer.  When the resistive layer obtains a sizable width, higher order corrections to the tearing stability parameter become important, and the mode grows less rapidly.

In addition to comparison against well-known scaling laws, we benchmark \jorek results for classical tearing modes against the linear stability code PHOENIX.
Panel (b) of FIG. \ref{figbench} compares results for the resistive layer width at zero viscosity produced using the \jorek code and from PHOENIX simulations, obtained from Figure 7.13 of \citet{haverkort2013magnetohydrodynamic} and from \citet{Haverkort2016}.  These codes show close agreement in the regime of moderate resistivity, diverging only mildly at high resistivity.  As part of this benchmarking effort, convergence of the width of the resistive layer based on the number of finite elements in the poloidal plane and the number of harmonics in the toroidal direction were carefully tested \citep[for further details, see][]{Haverkort2016}.

\begin{figure}
\resizebox{3.5in}{!}{\includegraphics{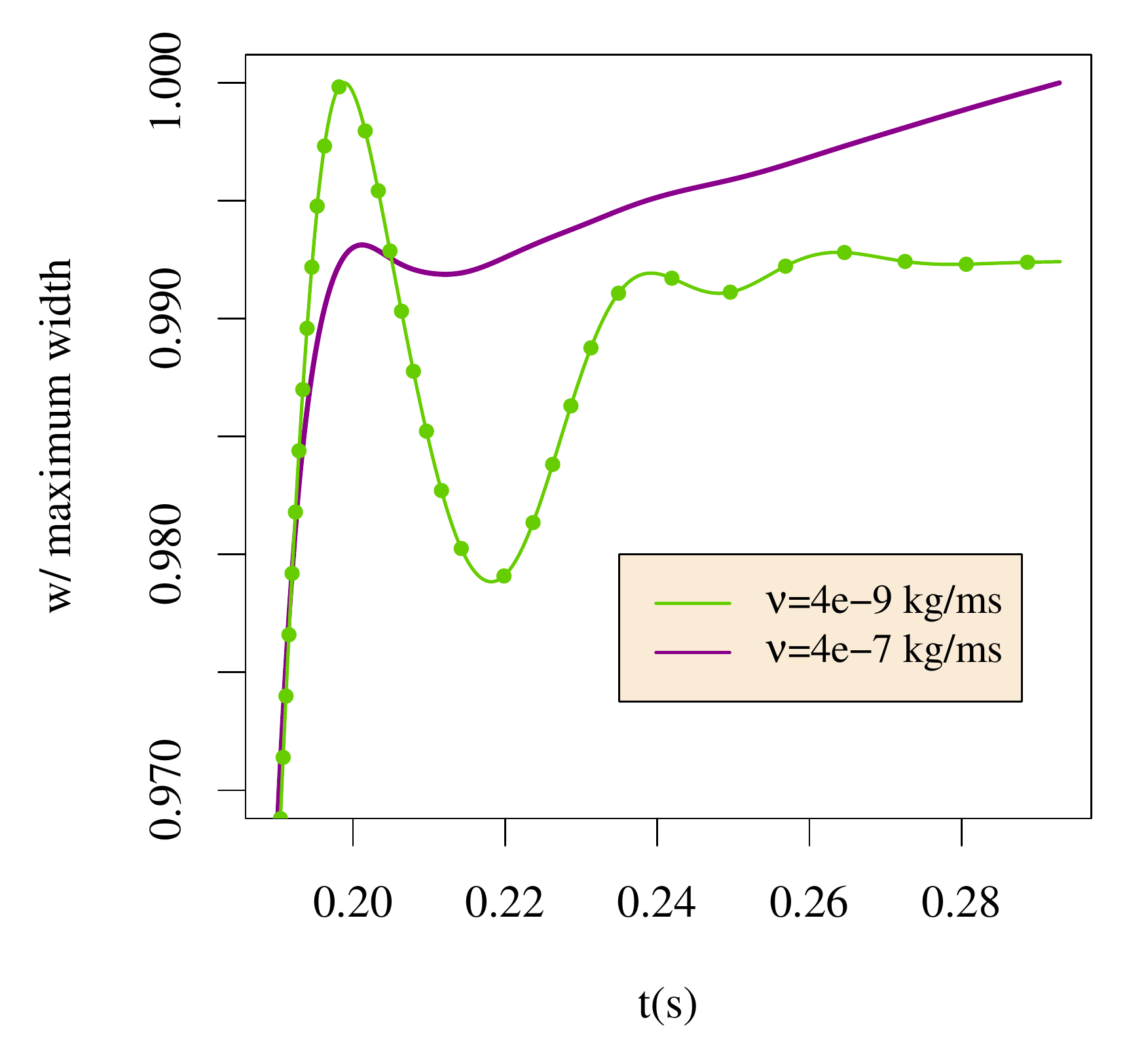}}
\caption{The width of the magnetic island $w$ vs time during the nonlinear saturation phase for simulations that are identical except for the magnitude of viscosity.   \label{figwithsat} }
\end{figure}

After the exponential growth phase the growth of the magnetic island width slows.  Typically a phase of linear growth, known as the Rutherford phase, is expected.  After the Rutherford phase, the island begins a process of nonlinear saturation \citep[e.g.][]{militello2004simple}.  For simulations with low viscosity, we observe that a shallow oscillation in the maximum width of the island occurs for some time.  A similar oscillation was also recently observed by \citet{poye2014saturation}.  This oscillation in magnetic island width is shown in FIG.~\ref{figwithsat}.

\section{Results \label{secresults}}

We perform a suite of 8 simulations that vary in toroidal resolution, represented by the number of toroidal harmonics $N$ used, and in the magnitude of the EC source, denoted by $\mathsf{max}(S_{\mathsf{EC}})$.  These simulations are all of an identical, large aspect-ratio circular tokamak with low plasma beta $\beta \sim  \mathcal{O} (10^{-4})$.  The viscosity $\nu =4 \cdot 10^{-9} \mathsf{kg/ms}$ used in these simulations is negligible so that it does not affect the tearing mode growth rate.  The resistivity $\eta = 2.5\cdot 10^{-6} \Omega$m is moderately higher than is expected by experiment, which is helpful for numerical efficiency, and not expected to impact our results.   
The Lundquist number is defined $S=v_{\mathsf{A}} \mathsf{L}/\eta$ where $v_{\mathsf{A}}$ is the Alfv\'en speed and $\mathsf{L}$ is a typical length scale, assumed to be the minor radius of the tokamak.  For the simulations considered in this work $S=2.8 \cdot 10^6$.  The collision frequencies $\nu_1$, $\nu_2$ are set to constants determined by comparison with a Fokker-Planck code, as described in \citet{westerhof2014closure}.  

To set the EC source so that it targets the resonant surface for the 2/1 tearing modes on both the inward and outward sides of the tokamak, we defined it as a Gaussian distribution centered at a fixed value of the poloidal magnetic flux $\psi$, normalized by its values at the axis of the poloidal plane and the outer boundary.  The standard deviation of this EC source is $0.08$ in units of the normalized poloidal magnetic flux; this is approximately equal to $0.08$m and remains constant throughout our simulations.  We find this width of the distribution to be well localized within our saturated magnetic island width. 

The parameters for the suite of 8 simulations are summarized in Table~\ref{simtab}.   In each simulation, the number of finite elements in the poloidal plane is 8281.  This is significantly higher poloidal resolution than is required to resolve the dynamics of the magnetic island; this higher number of finite elements is used to resolve the finer-scale dynamics of the EC current with high precision.  Toroidal modes $n=0,1,...,N$ are simulated. The lowest number of toroidal harmonics used in the simulations in Table~\ref{simtab} is $N= 3$.  During benchmarking, we found that the dynamics of the exponential growth phase of the magnetic island were well-resolved using $N \geq 3$.

\begin{table*}
\begin{center}
\caption{Classical 2/1 tearing mode simulation parameters. \footnote{For all simulations $\nu_1=9.47 \cdot 10^{3}$ Hz, $\nu_2=1.99 \cdot 10^{3}$ Hz, $a=1$m, $R=10$m, toroidal magnetic field $\vec{B}_{\phi} = 1.945$ T, and the number of finite elements in the poloidal plane is 8281.  The resistivity is $\eta = 2.5\cdot 10^{-6} \Omega$m and viscosity $\nu =4.\cdot 10^{-9} \mathsf{kg/ms}$.}  \label{simtab}}
\begin{tabular}{lccccccccccccc}
\hline\hline
Simulation                                                          & S1  & S2 & S3& P1  & P2 &  P3 & N1 &  T1    &  T2 &  T3         & F1 &  F2 &  F3 
\\ \hline
toroidal harmonics  $N$                                       &  3  &  3  &  3 &  6    & 6    &   6  &    9     & 13  & 13 & 13  & 21 & 21 & 21
\\
\hline
source $\mathsf{max}(S_{\mathsf{EC}}) (10^5 ~\mathsf{A} /\mathsf{m^2}\mathsf{s}) $         & 0  &  7.94 & 794   & 0  & 7.94 &    794  &   7.94    & 0  & 7.94 & 794     & 0  & 7.94 & 794
\\ \hline
\end{tabular}

\end{center}
\end{table*}

\subsection{Early evolution of the EC driven current}

\begin{figure}
\resizebox{3.5in}{!}{\includegraphics{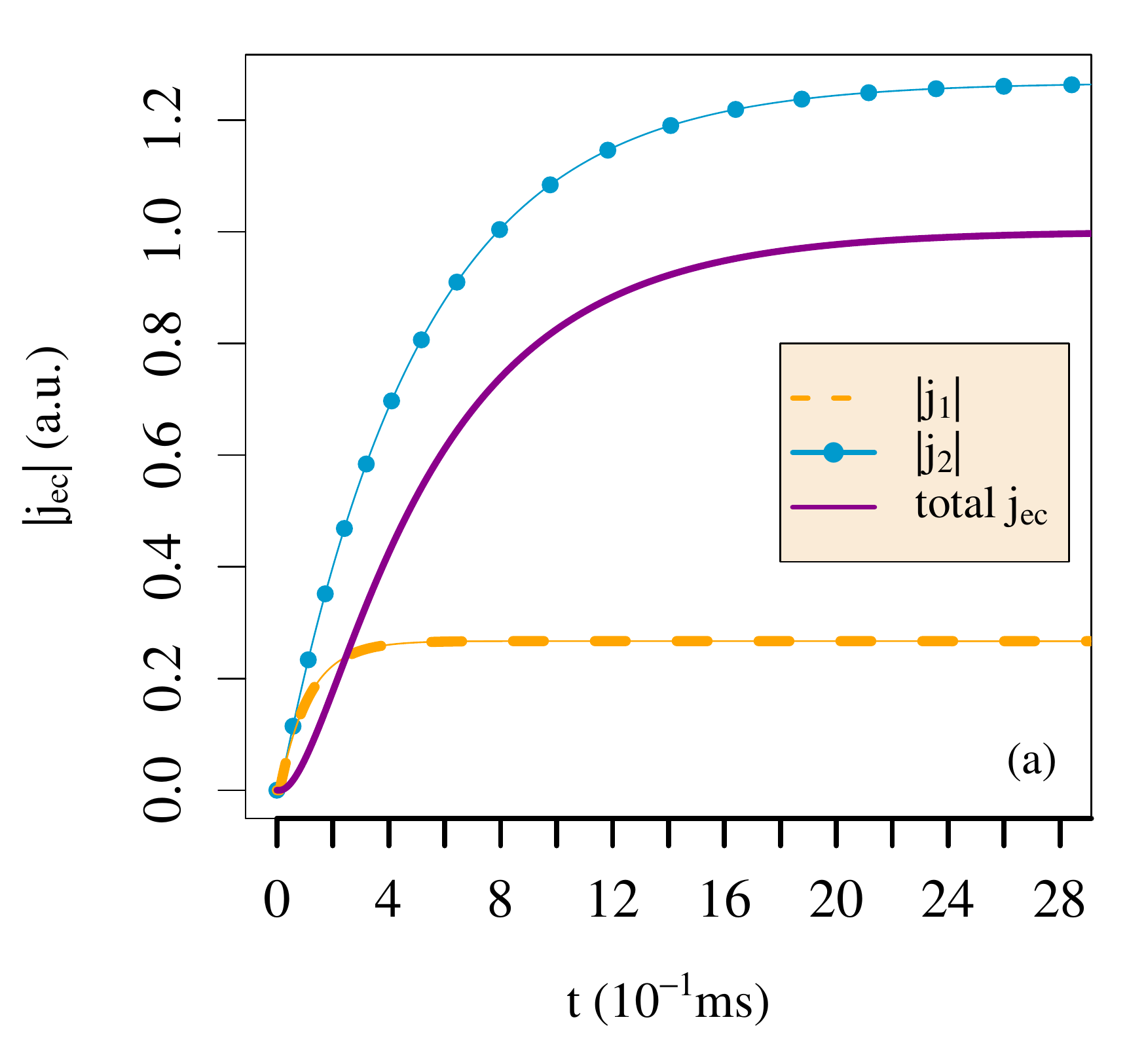}}\resizebox{3.5in}{!}{\includegraphics{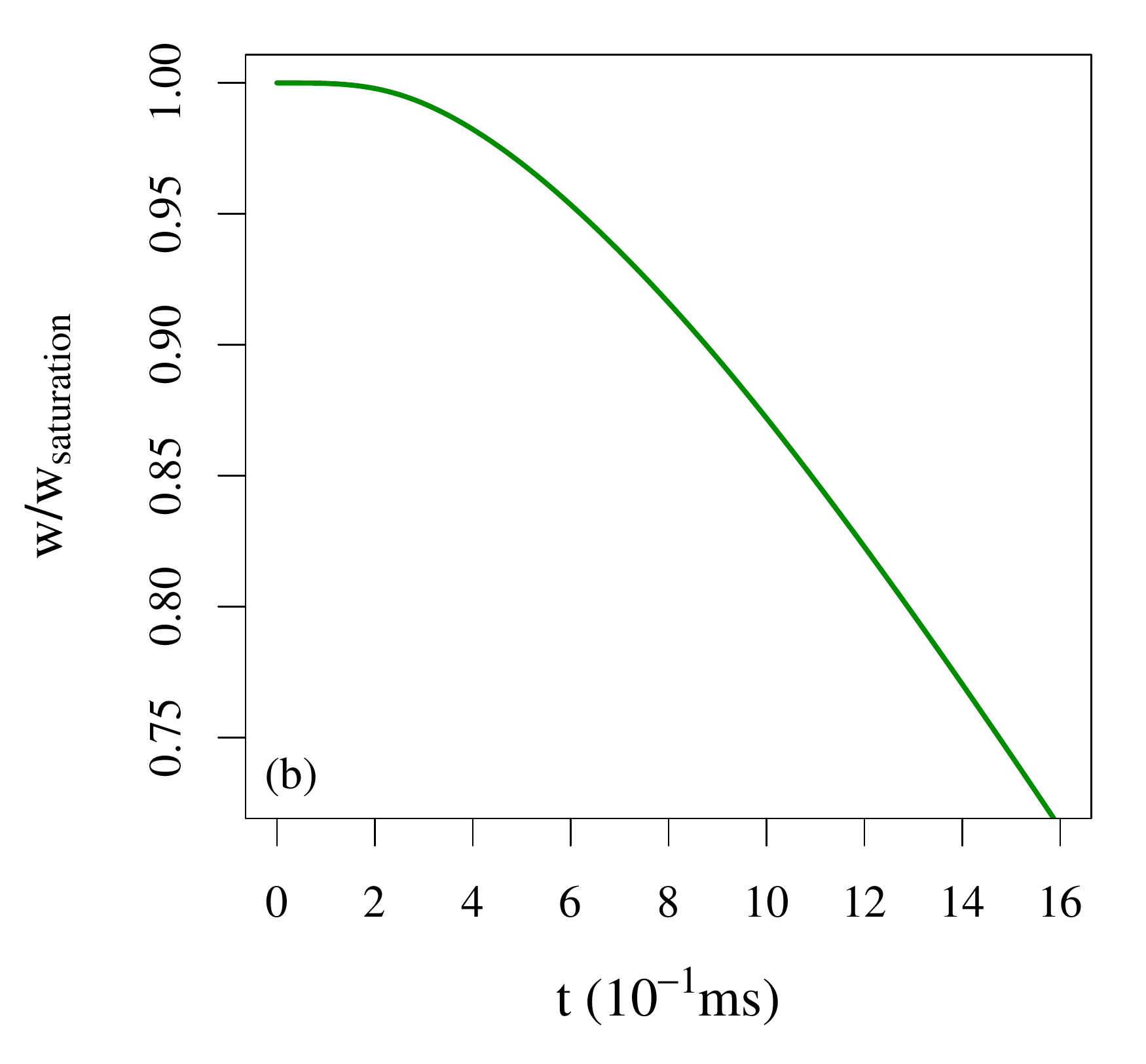}}
\caption{(a) Early-time evolution of the magnitude of the total EC current $j_{\mathsf{ec}}$ as well as the magnitude of its contributions $j_{1}$ and $j_{2}$. (b) The width of the magnetic island during the period where the EC current is relaxing toward a steady-state value.  Results are from simulation P3, described in Table~\ref{simtab}. \label{bench3} }
\end{figure}

Two distinct time scales exist in the evolution of the EC driven current, related to the two collision frequencies $\nu_1$ and $\nu_2$.  These two time scales can be observed in 
 the early time evolution of the total EC current $j_{\mathsf{ec}}$ in panel (a) of FIG.~\ref{bench3}.  The current perturbation that evolves faster, $j_1$, reaches a steady state on a time scale after approximately $0.2$ms.  For the more slowly evolving current perturbation, $j_2$, more than $1.4$ms is required for the current to reach a steady state.  Within $1.4$ms, the magnetic island width in simulation P3 is reduced to approximately 77\% of its width at saturation; the early evolution of the EC current can cover a significant period of time in our simulations. This is partly a reflection of the relatively high resistivity; at a resistivity realistic to experiment, these time scales are predicted to be more disparate. The drop in island width during suppression of the tearing mode is shown in panel (b) of FIG.~\ref{bench3}.  We note that in order to resolve this fast evolving current, a time step of approximately $2 \cdot 10^{-4}$ms is required.  The evolution of the total EC current does not change when significantly more resolution is used, either in the finite elements of the poloidal plane, or in spectral harmonics in the toroidal direction.    

In some early studies \citep[e.g.][]{yu1998stabilization} the EC driven current was assumed to be a function of the magnetic flux $\psi$.  
For the simulations described in Table~\ref{simtab}, the time for the EC driven current to spread along flux surfaces is approximately the time for it to be convected along the magnetic field lines; this time scale is faster than the collisional time scale on which the current reaches a steady state, so the steady-state EC driven current will tend to an equilibrium on the flux surfaces.  In the general case, the EC current will vary with the amplitude of the total magnetic field $B$ on the flux surfaces, and $j_{\mathsf{ec}}/B$ will be constant on the flux surfaces. 
However,  the present investigations, like earlier studies, have been performed in a large aspect-ratio tokamak; in this situation $j_{\mathsf{ec}}$ is  
independently expected to be approximately constant on the flux surfaces.
We note that in our simulations, an $n=0$ EC current source term creates an $n=1$ component of the EC current as the convection results in equilibration of $j_{\mathsf{ec}}$ over the flux surfaces inside the magnetic island.
  During early evolution when $j_{\mathsf{ec}}$ is growing and spreading along magnetic flux surfaces, however, it is not expected to be a function of the magnetic flux.

To examine how far $j_{\mathsf{ec}}$ is from a flux function, we define a measure  $\Sigma$, the flux-function error, such that
\begin{eqnarray}\label{eqsigmadef}
\Sigma = \langle \sigma_{\mathsf{jec}} / \bar{j}_{\mathsf{ec}} \rangle ~.
\end{eqnarray}
Here $\sigma_{\mathsf{jec}}$ is the standard deviation of $j_{\mathsf{ec}}$ along a surface of constant magnetic flux.  The average of the EC driven current along that surface is $\bar{j}_{\mathsf{ec}}$.  The brackets $\langle ... \rangle$ indicate an average of the ratio over all closed magnetic flux surfaces inside the magnetic island.  Thus $\Sigma$ measures in a global sense how far $j_{\mathsf{ec}}$ is from being a flux function.    At the separatrix, the field-line connection length approaches infinity; the contribution to $\Sigma$ should therefore be largest there.  We observe that the largest contributions to $\Sigma$ are from magnetic flux surfaces near the separatrix; interior to the magnetic island contributions to $\Sigma$ are smaller but exhibit no clear trend.

\begin{figure}
\resizebox{3.5in}{!}{\includegraphics{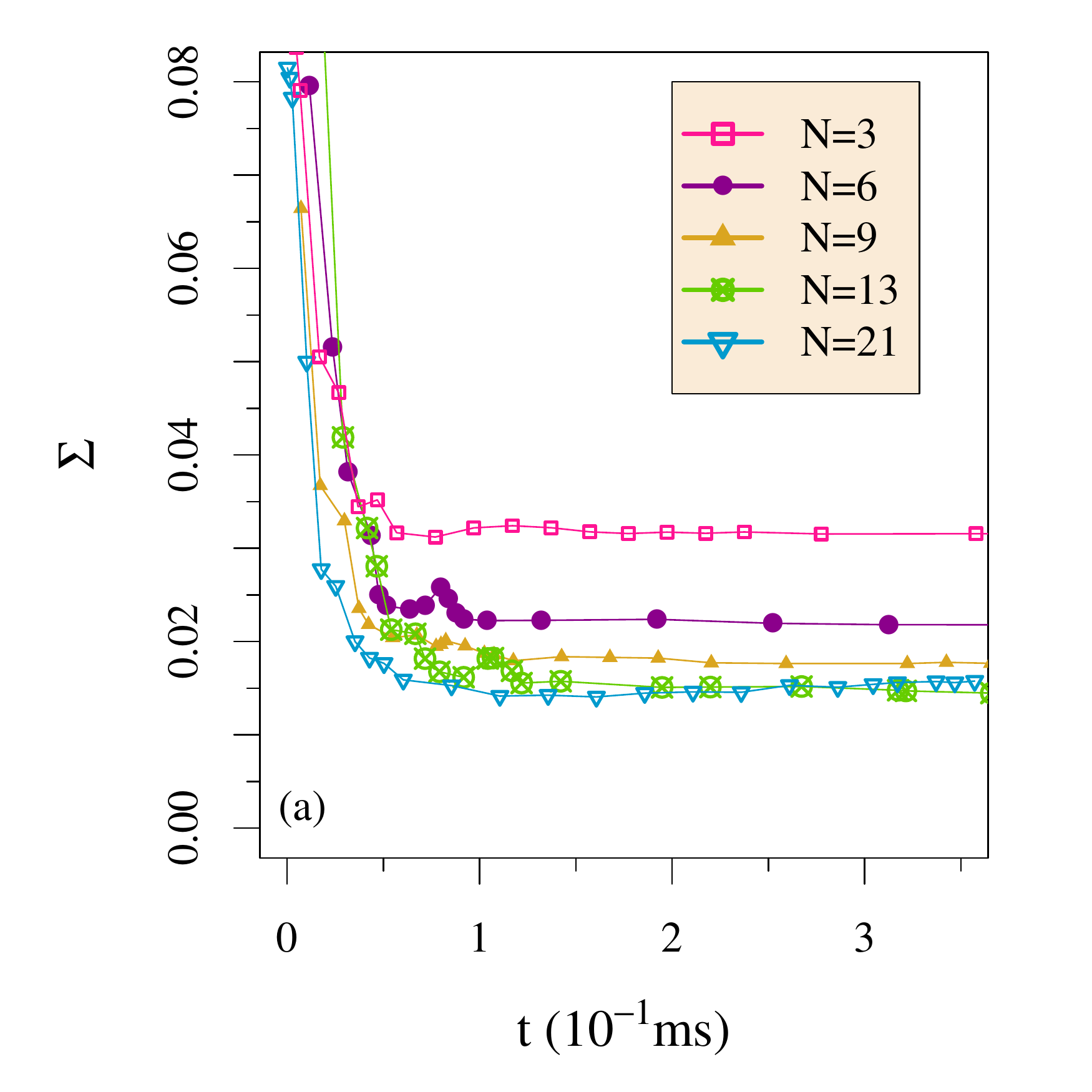}}\resizebox{3.5in}{!}{\includegraphics{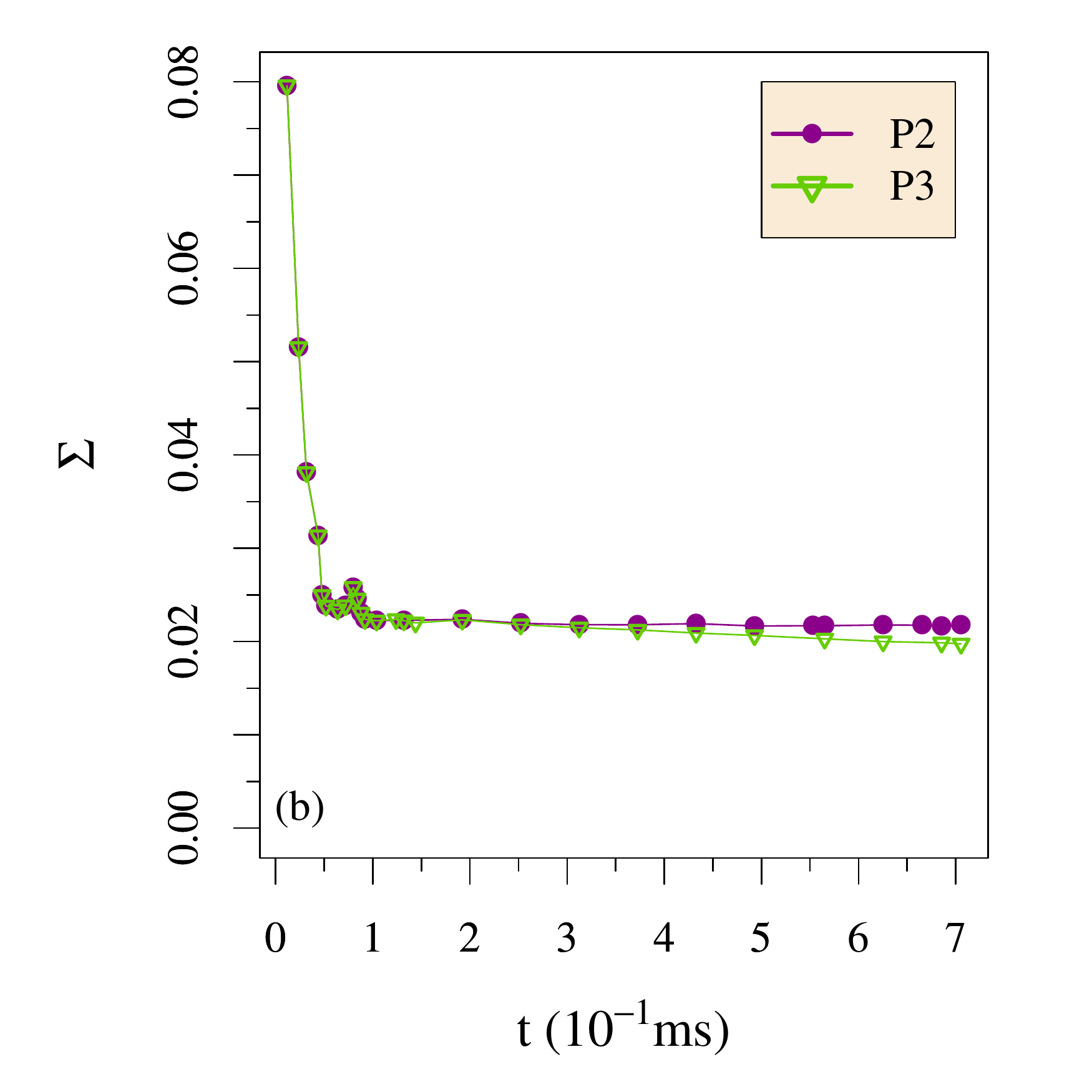}}
\caption{\label{figerrjec} 
The evolution of the flux-function error $\Sigma$, defined in eq.~\eqref{eqsigmadef}, for (a) simulations S2, P2, N1, T2, and F2 that are identical except for the number of toroidal harmonics used, and (b) simulation P2 and P3 that are identical except for the source amplitude for the EC driven current. }
\end{figure}

Panel (a) of FIG.~\ref{figerrjec} shows how $\Sigma$ evolves in time for simulations that use different numbers of toroidal harmonics $N$, but otherwise have identical parameters.  In each of these simulations, $\Sigma$ is initially high.  After approximately $0.2$ms,  $\Sigma$ has dropped to a steady, much lower level.  Before this steady level of flux-function error $\Sigma$ is reached, $\Sigma$ changes non-linearly, and sometimes non-monotonically, as the EC driven current spreads throughout the magnetic island.  The final level of $\Sigma$  is lower for simulations with a larger number of toroidal harmonics $N$.  This implies that once a steady state EC current is established it is close to a flux function, with some small constant error.  As the toroidal direction is better resolved, the approximation of the EC current as a function of the magnetic flux also becomes a better approximation.

For the simulations in panel (a) of FIG.~\ref{figerrjec} the source amplitude for the EC driven current, $\mathsf{max}(S_{\mathsf{EC}})$, was chosen to be 100 times smaller than a typical tokamak experiment.  For this source amplitude the magnetic island is not meaningfully affected by the application of the current.  Panel (b) of FIG.~\ref{figerrjec}
compares $\Sigma$ for simulations P2 and P3, which differ only in the source amplitude $\mathsf{max}(S_{\mathsf{EC}})$; simulation P2 uses the same low source amplitude, where simulation P3 uses a source amplitude one hundred times higher, in the range of a typical tokamak experiment.  
For simulation P2, a constant value of $\Sigma$ is reached.
In contrast, in simulation P3, $\Sigma$ continues to drop slowly as the magnetic island shrinks, and magnetic flux surfaces evolve under the influence of the applied current.

\subsection{Early suppression of the tearing mode}

\begin{figure}
\resizebox{3.5in}{!}{\includegraphics{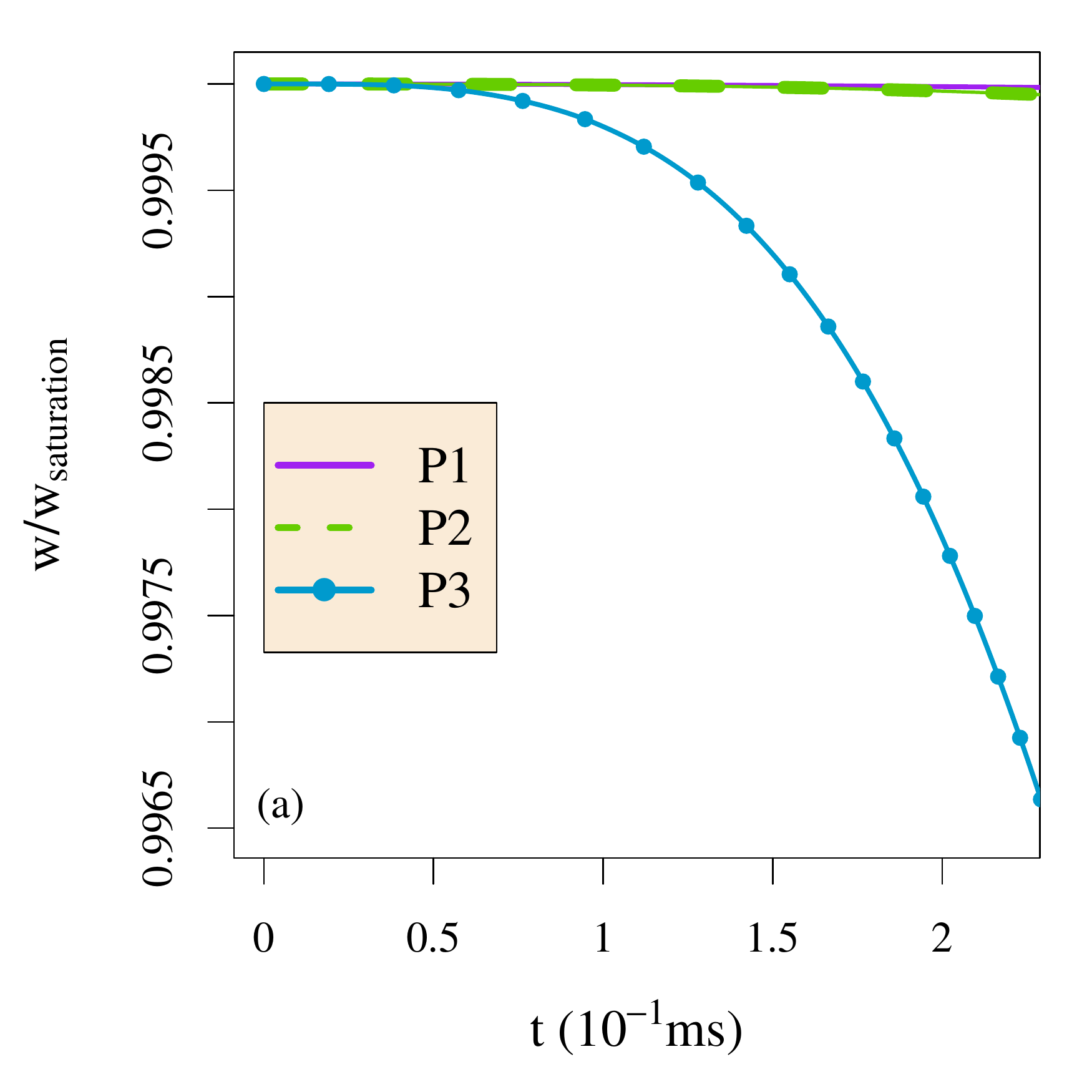}}
\resizebox{3.5in}{!}{\includegraphics{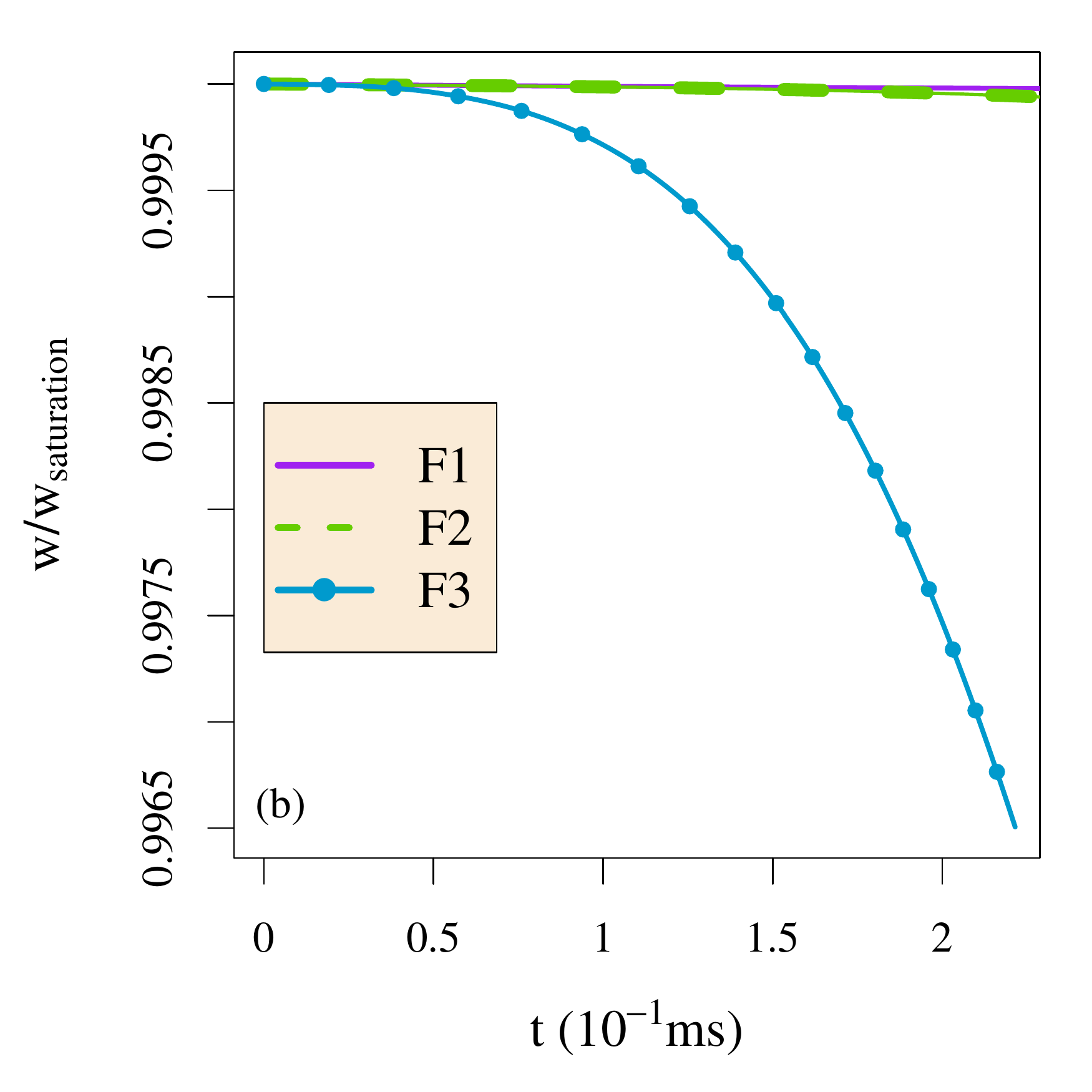}}
\caption{\label{figwidth} Width of the magnetic island during the period where the fast-evolving current perturbation $j_1$ is relaxing towards its steady-state value for (a) simulations with 6 toroidal harmonics: P1, P2, and P3, and (b) simulations with 21 toroidal harmonics: F1, F2, and F3.  Solid lines indicate simulations with no applied ECCD.  Dashed lines indicate simulations with small applied ECCD.  Dotted lines indicate simulations with high levels of EC driven current.}
\end{figure}

When ECCD is applied, the width $w$ of the magnetic island immediately begins to shrink proportionally to the magnitude of the current.  In FIG.~\ref{figwidth}, panels (a) and (b) follow the width of the magnetic island during the phase where the fast-evolving current perturbation $j_1$ is relaxing toward a steady-state, in two independent sets of simulations that use a moderate number of toroidal harmonics, and a high number of toroidal harmonics respectively.  In each panel, lines are drawn that show the evolution of island width for no ECCD (solid line), for a small amplitude of EC driven current (dashed line), and a driven current amplitude that is one hundred times larger (dotted line).  The local maximum value of the large EC current is 15\% of the local maximum of the full current; the total EC current applied is approximately 5\% of the total current.  For the small amplitude of EC driven current, the width of the island changes a negligibly small amount during the relaxation toward a steady-state EC current.  When we correct for the slightly different initial values of $dw/dt$ in the two sets of simulations represented in panels (a) and (b) of FIG.~\ref{figwidth}, the results exhibit a high degree of similarity.  At this level, the width of the magnetic island as it is suppressed appears to be independent of the number of toroidal harmonics $N$.

\begin{figure}
\resizebox{3.5in}{!}{\includegraphics{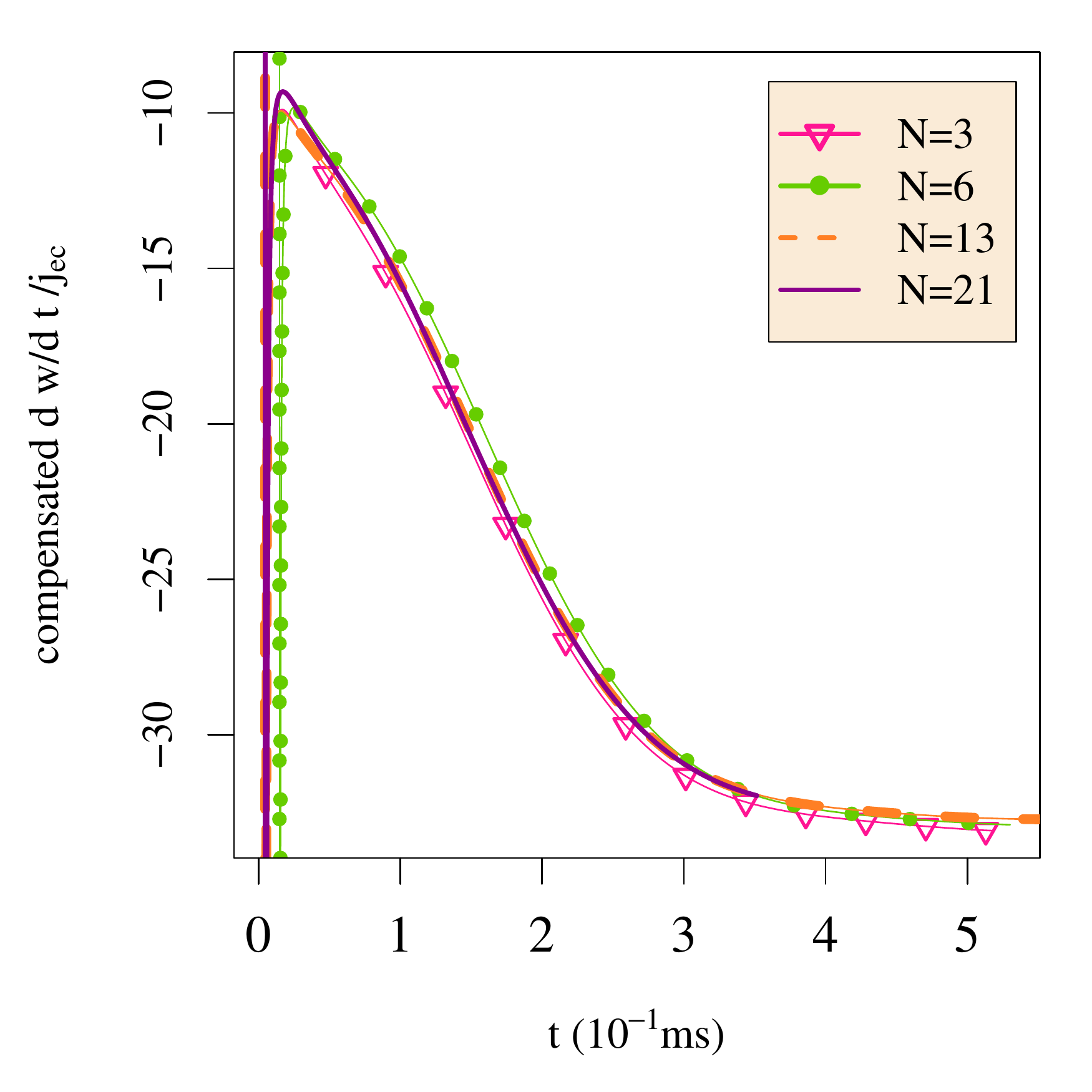}}
\caption{\label{figdercomp} The compensated time derivative of the island width in arbitrary units for simulations S3, P3, T3, and F3. }
\end{figure}

After a magnetic island has reached its maximum width, the island width can change by small amounts in time, shallowly growing or shrinking even before ECCD has been applied.  For simulation P1 the maximal amplitude of oscillation is 1\% of the saturated island width.  Therefore to compare the size of the island shrinking under the influence of ECCD in simulations P2 and P3 precisely, we need to consider the value of $dw/dt$ from simulation P1, an identical simulation with no ECCD.
We define the compensated time derivative of the island width:
\begin{eqnarray}\label{eqsigmadef}
\frac{1}{j_{\mathsf{ec}}(t)}\frac{d w}{dt} \biggr \rvert_{\mathsf{comp}} = \frac{1}{j_{\mathsf{ec}}(t)} \left(\frac{d w}{dt} \biggr \rvert_{\mathsf{with~ECCD}}-\frac{d w}{dt} \biggr \rvert_{\mathsf{no~ECCD}} \right)
\end{eqnarray}
By dividing by the total EC current, this definition of the compensated time derivative of the island width allows for comparison between simulations with different levels of applied EC driven current.
The compensated time derivative of the island width is plotted in  
FIG.~\ref{figdercomp} for four simulations that use increasing numbers of toroidal harmonics but otherwise have identical parameters.  Because the results in FIG.~\ref{figdercomp} are similar, we conclude that the suppression of the magnetic island is well-resolved in the toroidal direction for each of these simulations.

From FIG.~\ref{figdercomp} the multiple time scales of our physical problem are evident.
On the fastest time scale $\mathcal{O}(10^{-1} \mathsf{ms})$, $j_{\mathsf{ec}}$ spreads
over the magnetic flux surfaces (see also FIG. \ref{figerrjec}) and the fast evolving current perturbation $j_1$ reaches a steady state (see also FIG. \ref{bench3}). If the effect of the EC current on the island were entirely produced
by its helical component, as discussed in \citet{westerhof2016new},
 then the compensated time derivative of the island width would be constant after approximately $0.1$ms. 
However, in 
FIG.~\ref{figdercomp} we observe that the compensated time derivative of the island width
reaches a constant only on a longer time scale of $\sim 0.4$ms.  
This longer time scale can be related to a local EC current diffusion time.
 The local EC current diffusion time is defined
 \begin{eqnarray}
  \tau_{\mathsf{ECdiff}}=\mu_0 \sigma^2/\eta~,
 \end{eqnarray} 
  where $\sigma$ is the standard deviation of the Gaussian-shaped EC current source, and $\mu_0$ is the permeability of free space.  For all of the simulations in in Table~\ref{simtab}, we find $\tau_{\mathsf{ECdiff}} \approx 0.32$ms.   The evolution of the compensated time derivative of the island width and the local EC current diffusion thus act on similar time scales.
This suggests that a contribution from the poloidally-averaged  $n=0$ component of the EC current also acts through the mode stability.    To examine this effect in greater detail, we repeat simulations P2 and P3, allowing only the $n=0$ contribution of the EC current to contribute to the magnetic island evolution.  In these test simulations, we find that the island is suppressed on approximately the same time-scale as when all $n$ modes are allowed to affect the magnetic island evolution.  Therefore the poloidally-averaged $n=0$ component of the EC current significantly impacts the tearing mode suppression in our simulations by altering the current density profile and thereby the tearing mode stability parameter.  A discussion on the balance of the effect of the $n=0$ and $n=1$ components can be found in \citet{westerhof2016new}.

\section{Summary and Conclusions \label{secconcl}}

Using a two-equation fluid model for the EC driven current derived by \citet{westerhof2014closure}, and a 3D reduced-MHD fluid model in the \jorek code, we have numerically shown that the steady-state EC driven current $j_{\mathsf{ec}}$ is approximately a function of the magnetic flux in a large aspect-ratio tokamak.  High resolution in the toroidal direction is required to accurately produce this applied current as a flux function.  When a lower resolution in the toroidal direction is used, the applied current varies from a flux function by an error that we find is typically reasonably small for $N \geq 6$, and decreases as $N$ is increased.  For $N = 6$, when a higher amplitude of EC driven current is produced, the steady-state $j_{\mathsf{ec}}$ is closer to a flux function.  

Although how evenly the steady-state EC driven current spreads along surfaces of constant magnetic flux is sensitive to 
the toroidal resolution, any error in the form of the applied EC current  due to the toroidal resolution
appears to have negligible impact on the size of the island during the early period of suppression of the tearing mode.

An interesting aspect of the two-equation fluid model that we use for the EC driven current is that it captures the early evolution of the EC driven current with better physical accuracy than previously-studied single equation models. The full early evolution phase, before the total EC driven current $j_{\mathsf{ec}}$ reaches a steady-state, corresponds to a drop in magnetic island width of approximately a quarter relative to its saturated size. Before the $j_1$ current perturbation reaches a steady-state, the EC driven current is not a function of the magnetic flux, nor is it theoretically expected to be.  We find that the applied current spreads along magnetic flux surfaces within the island on the time scale of the relaxation of the faster-evolving current contribution $j_1$.  We observe that the compensated time derivative of the island width, which we define to compare tearing mode suppression in different simulations, evolves over the EC current diffusion time scale, which is a longer time scale than for the relaxation of $j_1$.   Using targeted test simulations, we conclude that the suppression of the magnetic island is accomplished primarily by the $n=0$ component, with smaller contributions from higher $n$ components of the EC current.

Due to our two-equation fluid model, the early evolution of the EC current differs from, and is more accurate than, those produced by either a simpler one-equation fluid model or a fluid model that includes only diffusive terms \citep[e.g. as investigated by][]{fevrier2016first}.   However, in the simulations presented in this work, the stabilization of the magnetic island measured by the drop in island width proved insensitive to the form of the EC current during this early phase.  We thus expect that a simple diffusive model for the EC current would have a similar effect on island stabilization in the same setting.
Our results apply for a continuous waveform (CW) EC source that is localized only in the radial direction of the poloidal plane.  Future investigations that use a modulated EC source, localized both in the poloidal and toroidal directions, may produce differences.  Investigations of X-point and finite-aspect-ratio geometries are also planned.

\begin{acknowledgements}\small
Many thanks to Marina B\'ecoulet, Matthias H\"olzl, J.W. Haverkort and the participants of the ASTER project and the \jorek collaboration.  This work was performed on the Helios system at the Computational Situational Centre, International Fusion Energy Research Centre (IFERC-CSC), Rokasho-Japan under the grant NTMJOREK and the Cartesius system, the Dutch national supercomputer, at SURFsara, Amsterdam, Netherlands.  

This project was carried out with financial support from NWO. The work has been
carried out within the framework of the EUROfusion Consortium and has received
funding from the Euratom research and training programme 2014-2018 under grant
agreement No 633053. The views and opinions expressed herein do not necessarily
reflect those of the European Commission.

The research leading to these results has also received funding from the European Research Council under the European Union's Seventh Framework (FP7/2007-2013)/ERC grant agreement no. 320478.
\end{acknowledgements}

\bibliographystyle{apsrev}
\bibliography{jorekbib}

\end{document}